\documentclass{emulateapj}

\slugcomment{Submitted to ApJS}

\shorttitle{The morphological content and enviromental dependence of
the galaxy color-magnitude diagram at $z\sim0.7$}
\shortauthors{P. Cassata, L. Guzzo, A. Franceschini, et al.}

\begin{document}

\title{The Cosmic Evolution Survey (COSMOS): The morphological content
and enviromental dependence of the galaxy color-magnitude  
relation at $z\sim0.7$}

\author{
P. Cassata\altaffilmark{1,2},
L. Guzzo\altaffilmark{3,11}, 
A. Franceschini\altaffilmark{2},
N. Scoville\altaffilmark{4,5}, 
P. Capak\altaffilmark{4},
R. S. Ellis\altaffilmark{4},
A. Koekemoer\altaffilmark{6},
H. J. McCracken\altaffilmark{7},
B. Mobasher\altaffilmark{6},
A. Renzini\altaffilmark{8},
E. Ricciardelli\altaffilmark{2},
M. Scodeggio\altaffilmark{1}
Y. Taniguchi\altaffilmark{9},
D. Thompson\altaffilmark{10,4}
} 

\altaffiltext{$\star$}{Based on observations with the
NASA/ESA {\em Hubble Space Telescope}, obtained at the Space Telescope
Science Institute, which is operated by AURA Inc, under NASA contract
NAS 5-26555; also based on data collected at : the Subaru Telescope,
which is operated by the National Astronomical Observatory of Japan;
the XMM-Newton, an ESA science mission with instruments and
contributions directly funded by ESA Member States and NASA; the
European Southern Observatory, Chile; Kitt Peak National Observatory,
Cerro Tololo Inter-American Observatory, and the National Optical
Astronomy Observatory, which are operated by the Association of
Universities for Research in Astronomy, Inc. (AURA) under cooperative
agreement with the National Science Foundation; the National Radio
Astronomy Observatory which is a facility of the National Science
Foundation operated under cooperative agreement by Associated
Universities, Inc ; and the Canada-France-Hawaii Telescope operated by
the National Research Council of Canada, the Centre National de la
Recherche Scientifique de France and the University of Hawaii. }

\altaffiltext{1}{INAF-IASF Milano, via Bassini 15, I-20133 Milano,
Italy}
\altaffiltext{2}{Dipartimento di Astronomia, Universit\`a di Padova, Vicolo dell'Osservatorio 2, I-35122, Padova, Italy}
\altaffiltext{3}{INAF-Osservatorio Astronomico di Brera, via Bianchi
  46, I-23807, Merate (LC), Italy}
\altaffiltext{4}{California Institute of Technology, MC 105-24, 1200 East
California Boulevard, Pasadena, CA 91125}
\altaffiltext{5}{Visiting Astronomer, Univ. Hawaii, 2680 Woodlawn Dr., Honolulu, HI, 96822}
\altaffiltext{6}{Space Telescope Science Institute, 3700 SanMartin
Drive, Baltimore, MD 21218}
\altaffiltext{7}{Institut d'Astrophysique de Paris, UMR7095 CNRS, Universit\'e Pierre et Marie Curie, 98 bis Boulevard Arago, 75014 Paris, France}
\altaffiltext{8}{INAF - Osservatorio Astronomico di Padova, vicolo dell'Osservatorio 5, I-35122 Padova, Italy}
\altaffiltext{9}{Astronomical Institute, Graduate School of Science,
         Tohoku University, Aramaki, Aoba, Sendai 980-8578, Japan}
\altaffiltext{10}{Large  Binocular Telescope Observatory, 933 N. Cherry Ave.,  Tucson, AZ 85721-0065}
\altaffiltext{11}{Visiting Scientist, ESO and MPA/MPE, Garching, Germany}

\begin{abstract}
We study the enviromental dependence and the morphological composition
of the galaxy color-magnitude diagram at $z\sim 0.7$, using a pilot
sub-sample of the COSMOS survey.  The sample includes $\sim 2000$
galaxies with $I_{AB}<24$ and photometric redshift within
$0.61<z<0.85$, covering an area of 270 square-arcmin.  Galaxy
morphologies are estimated via a non-parametric automatic technique.
The $(V-z')$ vs. $z'$ color-magnitude diagram shows a clear
red-sequence dominated by early-type galaxies and also a remarkably
well-defined ``blue sequence'' described by late-type objects.  While
the percentage of objects populating the two sequences is a function of
environment, following a clear morphology/color-density relation also
at this redshift, we establish that their normalization and slope are
independent of local density.  We identify and study a number of
objects with ``anomalous'' colors, given their morphology, polluting
the two sequences.  Red late-type galaxies are found to be mostly
highly-inclined or edge-on spiral galaxies, for which colors are
dominated by internal reddening by dust.  In a sample of
color-selected red galaxies, these would represent a 33\%
contamination with respect to truly passive spheroidals.  Conversely,
the population of blue early-type galaxies is composed by objects of
moderate luminosity and mass, concurring to only $\sim5\%$ of the mass
in spheroidal galaxies. The majority of them ($\sim70\%$) occupy a
position in the $\mu_B$-$r_{50}$ plane not consistent with them being  
precursors of current epoch elliptical galaxies.  Their
fraction with respect to the whole galaxy population does not depend
on the environment, at variance with the general early-type class.  In
a color-mass diagram, color sequences are even better defined, with
red galaxies covering in general a wider range of masses at nearly
constant color, and blue galaxies showing a more pronounced dependence
of color on mass.  While the red sequence is adequately reproduced by
models of passive evolution, the blue sequence is better interpreted
as a specific star-formation sequence.  The substantial invariance of
its slope and normalization with respect to local density suggests
that the overall, ``secular'' star formation is driven more by galaxy
mass than by environment.

\end{abstract}


\section{Introduction}

It has become clear in recent years that galaxy colors show a bimodal
distribution (Strateva~et~al.~2001, Hogg~et~al.~2002,
Blanton~et~al.~2003): in a rest-frame color-magnitude diagram galaxies
tend to segregate between a ``red sequence'' (similar to, but less
tight than that observed for cluster galaxies) and a ``blue cloud''.

The red sequence is mostly composed by spheroidal galaxies, a class of
objects of special interest as it includes the most massive galaxies,
adding up to represent about half the stellar mass in the Universe
(Bell~et~al.~2004a,b; Baldry~et~al.~2004, Hogg~et~al.~2004).  This
behaviour seems to be already established at fairly high
redshifts (Bell~et~al.~2004a; Giallongo~et~al.~2005;
Cucciati~et~al.~2006; Franzetti~et~al.~2006).  The SDSS data
(Hogg~et~al.~2004) show that the blue cloud contains mainly late
(spiral) morphological types, while the bulk of the red sequence
consists of early-type galaxies. Hogg~et~al.~(2004) find also that
galaxies with high S\'ersic indices (indicating an early-type
morphology) dominate the red sequence, almost independently of local
density. Balogh~et~al.~(2004), using a color criterion to describe
galaxy populations, also find that the colors of the red and the blue
populations depend weakly on the environment.

A dependence of galaxy colors on the environment would seem to be a
natural expectation of a hierarchical scenario: the most dense regions
are also those that collapse first, and therefore those appearing as
most evolved at any epoch.  If the baryonic component simply follows
the evolution of dark matter haloes, high density regions should
contain galaxies that are older and redder than the low-density field
population. How these properties of the stellar population are
inter-laced with the construction of the morphological type and its
relation to the environment is not obvious, given the impossibility of
semi-analytical models to predict galaxy morphologies, a part from
adopting simple recipes.  Thus, attempts to explain the observed
bi-modality in the rest-frame color-magnitude distribution seem to be
succesfull in reproducing the gross color features of the population
(Menci~et~al.~2005; Dekel~et~al.~2005), but obviously do not tell us
much on the specific morphological differentiation.

Investigation of the actual morphological composition of the red
sequence at high redshift has been performed so far by
Bell~et~al.~(2004b), using the GEMS survey, and by
Weiner~et~al.~(2005), using the smaller DEEP-1 survey. Both works
consistently show that the red sequence also contains galaxies of type
later than Sa for a percentage between 20\% and 25\%.  They do not
explore, however, the dependence of the morphological composition of
the color sequences on the environment.

The Cosmic Evolution Survey (COSMOS, Scoville~et~al.~2007a) provides
us with an ideal combination of area, high-resolution HST imaging and
multi-band information to perform such study.  In this paper we
present a first pilot study based on a sub-sample of the COSMOS
catalogue, with mean redshift $z\simeq 0.7$.  This sample is centered
around a large-scale structure discovered by an adaptive-filter search
in redshift space (Scoville~et~al.~2007b). A parallel analysis
(Guzzo~et~al.~2007, Paper I hereafter) investigates in detail the
local environment of galaxies in this sample using projected densities
and X-ray and weak-lensing measurements. A comparison with the total
parent COSMOS sample within the same redshift range shows this
subsample to be representative with respect to galaxy color,
luminosity and mass. While we are already working on extending our
analysis to the whole area using improved spectroscopic and
photometric redshifts, the current uncertainties on galaxy distances
and consequently on measured local densities have suggested us this
first exploratory work in an area where these quantities are better
under control (see Paper I).

The paper is organized as follows: in \S~\ref{data} we give a brief
description of the data used here; in \S~\ref{env_col_mor} we present
the sample selection, we review the techniques used to estimate local
densities and we present the tool used to assign morphological types;
in \S~\ref{cmr} we describe the color-magnitude relation, and its
morphological content; in \S~\ref{A1} we analyze in more detail the
properties of red late-type and blue early-type populations; in
\S~\ref{cMassr}, we introduce the color-mass relation with the aim of
understanding the color-magnitude diagram; we discuss these findings
in \S~\ref{discussion} and summarize paper results in \S~\ref{end}.

We adopt throughout the paper a ``concordance'' cosmological model,
with $H_o=70$ km s$^{-1}$ Mpc$^{-1}$, $\Omega_M=0.3$, $\Omega_\Lambda
= 0.7$. However, when needed, we quote lengths and densities in units
of $h=H_0/100$.

\section{The Data}
\label{data}

The COSMOS {\it Treasury} project is centered upon a complete 2-degree
survey in the near-infrared ($\sim I+z$) F814W band using the Advanced
Camera for Surveys (ACS) on board HST.  The field is centered at
$\alpha$(J2000) = $10^{\rm h} ~ 00^{\rm m} ~ 28.6^{\rm s}$ and
$\delta$(J2000) = $+02^\circ ~ 12' ~ 21.0''$ (Scoville~et~al.~2007a,
2007c).  The ACS camera, with a field of view of 203 arcsec side, has
covered the whole field with a mosaic of 590 tiles, corresponding to
one orbit each, split into 4 exposures of 507 seconds dithered in a
4-point box pattern.  The final 590 images, obtained with the
``Multi-Drizzled'' software (Koekemoer~et~al.~2002) using the latest
improved geometric distortion corrections for ACS, are
$5600\times5600$ pixels each, with 0.05 arcsec pixels and an absolute
astrometric accuracy of better than 0.1 arcsec.  More details and a
full description of the ACS data processing and products are provided
in Koekemoer~et~al.~(2007).

The COSMOS 2-square degree field has been also fully covered by
optical ground-based observations (Taniguchi~et~al.~2007), using
Suprime-Cam on the 8.2~m Subaru Telescope on Mauna Kea
(Kaifu~et~al.~2000).  The Suprime-Cam panoramic camera consists of
$5\times 2$ CCDs of 2k $\times$ 4k pixels, with a pixel scale of $0.2$
arcsec pixel$^{-1}$ (Miyazaki~et~al.~2002).  During two observing runs
in January and February 2004, $B$, $V$, $r'$, $i'$, and $z'$ band
images of the whole COSMOS field were obtained.  These data were
complemented with further deep imaging in $U$ and $i$ from CFHT, and
in $K$ from KPNO/CTIO, UH88 and UKIRT.  The $K$-band data are
essentially complete to $K_{AB}=20.8$. The detection of the
sources has been performed on a combination of the Subaru $i'$ and
CHFT $i$ original PSF images. The photometry, on the other hand, has
been measured within 3'' apertures in PSF omogenized images. In
pratice, each image is smoothed to achieve the same $fwhm$ of the
image with the largest PSF (the $K$-band), and then the photometry is
measured in a 3'' aperture. This procedure reduces the effects of the
PSF variation from band to band and has the advantage that the total
correction factors are identical in all bands. The resulting global
photometric catalogue contains around 440,000 objects to
$I_{AB}=25$. More details on the observations, data reduction {and
assembly of the catalogue} are given in Taniguchi~et~al.~(2007) and
Capak~et~al.~(2007a).

Photometric redshifts were derived from this catalogue using a
Bayesian Photometric Redshift method (BPZ) (Benitez~2000;
Mobasher~et~al.~2007), which uses six basic SED types, together with a
loose prior distribution for galaxy magnitudes.  The output of the
code includes, in addition to the best estimate of the redshift, the
68 and 95\% confidence intervals, the galaxy SED type, its absolute
magnitude in several bands, and stellar masses.  Based on nearly 1200
objects for which spectroscopic observations are available in the
COSMOS area (z-COSMOS, Lilly~et~al.~2007), the difference between
spectroscopic and photometric redshifts has a typical {\it rms} value
$\sigma_z\simeq 0.03(1+z)$ (not considering catastrophic
failures).  The quality of the photometric redshift however does
not strongly depend on the SED type.  As discussed in detail in
Mobasher et al. (2007), we have $\sigma_z=0.034(1+z)$ for early-type
(E/S0) galaxies, $0.030(1+z)$ for spiral galaxies and $0.042(1+z)$ for
starbursts, with a number of catastrophic failures below 2\%. These
figures are consistent with the restricted statistical sample of
spectroscopic redshifts available within the large-scale structure at
$\left<z\right>=0.73$ studied here (Paper I).  To improve typical
redshift errors, for this analysis we shall restrict the photometric
redshift sample to $I_{AB}<24$.  Stellar masses were estimated from
the same SED fits and rely, in particular, on the available K-band
photometry.  Indeed, at the moderate redshifts of our sample galaxies,
the K-band effectively probes the low-mass stellar content dominating
the galactic mass budget (while higher-z determinations will need the
longer-wavelength Spitzer photometry). More detailed information on
the quality of the photometric redshifts and stellar mass estimates
can be found in Mobasher~et~al.~(2007).

\section{SAMPLE DEFINITION AND DATA ANALYSIS}
\label{env_col_mor}

Scoville~et~al.~(2007b) use an adaptive-smoothing algorithm to search 
for large-scale structures within the COSMOS field, by exploiting
photometric redshift to divide the survey volume in ``slices''.
This analysis provides evidence -- among a number of systems at 
different redshifts -- for a conspicuous
overdensity peaking in the redshift interval $z=[0.65,0.85]$. 
The reality of this structure, with a mean redshift
$\left< z\right> =0.73$, has been spectroscopically confirmed by early
redshift measurements of 14 galaxies in this area (Comastri et
al., priv. comm.) and, more recently, by the first data from the VIMOS 
spectroscopic survey of the field (Lilly~et~al.~2007).  
The higher-density peaks within these structures are detected both 
in the XMM X-ray data (Finoguenov~et~al.~2007) and in the weak-lensing
mass-reconstruction map (Massey~et~al.~2007; see Paper I for comparisons).

We concentrate here (as in Paper I) on a sample of galaxies selected
from an area of $18\times15$ arcmin$^2$ that roughly encloses the
main part of the detected structure.  This area is covered by 30
ACS tiles of the overall COSMOS ACS survey.

%
\begin{figure}
\epsscale{1.}
\plotone{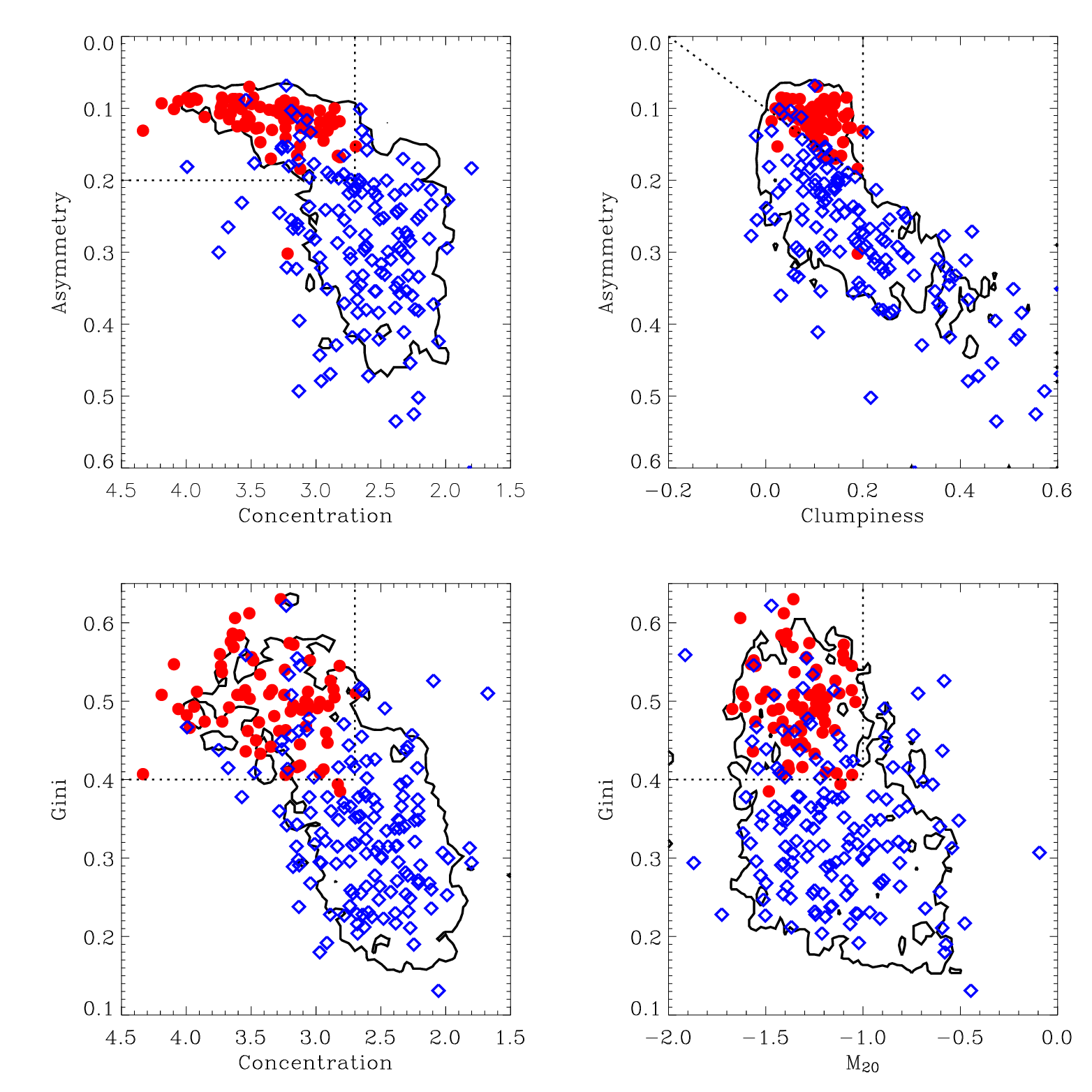}
\caption{Galaxies that have been visually classified as early-
(filled circles) and late-type galaxies (open lozenges) in our control
sample, as they appear in the multi-dimensional space defined by
Concentration, Asymmetry, Clumpiness, Gini and M20. The contours in
each panel indicate the region enclosing 90\% of the sample. The
dotted lines indicate the boundaries defined by eq. \ref{eqCAS}. Note
that galaxies with ``wrong'' morphological classification appearing in
the 2-parameter plots, are in most cases the result of projection,
being excluded when the whole 5-dimensional selection box is
considered. }
\label{casgm}
\end{figure}
%
Following Dressler~(1980), galaxy projected densities are estimated as
$\Sigma_{10}= {10 / \pi d_{10}^2} $, where $d_{10}$ is the projected
comoving distance (in h$^{-1}$ Mpc) to the $10^{th}$ nearest galaxy,
and then corrected for background contamination.  The robustness of
our local density estimates has been extensively tested using
simulations (Paper~I, Appendix). In particular, we have applied our
technique to a mock catalog generated from the Millennium simulation
(Kitzbichler~\&~White 2006), and we have showed that projected
densities correlate well with real 3-d ones above
$\Sigma_{10}=10\div20Mpc^{-2}h^2$. Capak~et~al.~(2007), using a
different technique, find similar results. As thoroughly discussed in
Paper I, we find that our estimate is fairly robust with respect to
varying the redshift slice thickness, and on this basis adopt as
standard for our analysis the redshift interval $\Delta z =
[0.61,0.85]$ corresponding roughly to $\pm 2\sigma_z$ around the
structure mean redshift.  The sample includes 2041 galaxies to
$I_{AB}=24$.

\subsection{Morphological classification}\label{mor_cla}

The size of the COSMOS sample naturally suggests the use of an
automatic and objective morphological classification technique.
Our aim in this paper is to separate galaxies into early-types
(including ellipticals and S0) and late types (spirals of any B/D,
irregulars and mergers). We have therefore based our
morphological estimates on non-parametric measurements of the galaxy
light distribution, extending the work of Cassata~et~al.~(2005).  We
have made use of a parameter set including Concentration $C$,
Asymmetry $A$, Clumpiness $S$, Gini index $G$ and Moment $M20$
(Conselice~2003, Abraham~et~al.~2003, Lotz~et~al.~2004). We estimated
these parameters for all galaxies in the 30 ACS/HST images covering
the $\left <z\right> =0.73$ structure.  At this redshift, the observed
F814W band approximately corresponds to rest-frame $B$ band. The
thickness of our redshift slice is small enough to ensure that
morphological K-correction and cosmological dimming effects can be
neglected inside it.  Most of the results discussed here are based on
internal comparisons (e.g. for different density regimes); when
reference to local (e.g. SDSS) morphological types is made, these
effects are in general small for our two simple classes of early- and
late-type galaxies (see Brinchmann~et~al.~1998) and in any case well
within the differences in morphological definitions among different
authors.

The technique used is directly derived from Cassata~et~al.~(2005), to
which we refer for all technical details about the methods for
measuring Concentration, Asymmetry and Clumpiness.  In this paper we
have added two further estimators, namely the {\it Gini index} and
{\it M20} parameter. 

The Gini coefficient is a concentration parameter, measuring how
fairly the light is distribuited among the galaxy pixels. In pratice,
it is high when the light is concentrated in a small number of pixels,
wherever they are located inside the galaxy. On the contrary, it is
low when the galaxy light is spread over a large number of pixels.
M20 is the moment of the brightest $20\%$ of galaxy flux: it is
another concentration parameter, that however is more sensitive than
Concentration to any off-axis light clump. Thus, it correlates with
Concentration for normal galaxies, and is expected to diverge from
that for non-symmetric objects with clumps of light far from the
center.  Parameters are defined and measured as in
Lotz~et~al.~(2004). The only difference is in the segmentation
technique used to assign image pixels to galaxies. In this work, we
select an area around each galaxy 4 times larger than the isphotal
area measured by SExtractor, centered on the galaxy centroid, and then
consider only pixels brighter than 1.5 times the rms of the
background. In fact, given that we are dealing here with galaxies
almost at the same redshift, we do not need to use Petrosian
apertures, that have the advantage of including the same part of
galaxy image independently of redshift, minimizing surface
brightness dimming effects.  This latter approach is being applied in
the extension of our work to the whole COSMOS area and to different
redshifts, which is in preparation.

Stars were removed from the sample using {\it SExtractor} (Bertin \& Arnouts
1996), according to their CLASS-STAR and FLUX-BEST parameters measured
on the ACS data, for which this method proves to be reliable (much
more than on ground-based images).   

As shown by Cassata~et~al.~(2005), there is a precise locus in the CAS
parameter space where early-type galaxies typically fall. To re-define
this region for the COSMOS ACS data, adding the new two
parameters, we used a control sample of 211 galaxies, extracted from
two ACS/HST tiles including the core of the main structure, for
which we performed a visual classification into early-types
(ellipticals and S0) and late-types (spirals, irregulars
and mergers).  In Figure~\ref{casgm} we show the distribution of these
galaxies in the space defined by Concentration, Asymmetry, Clumpiness,
Gini and M20 parameters.
Using the following empirical selection rules:
\begin{eqnarray}
  C > 2.7 \hspace{2mm} \& \hspace{2mm} A < 0.2  \hspace{2mm} \&
  \hspace{2mm} 2A-S < 0.2 \hspace{2mm} \& \hspace{2mm} S < 0.27 \hspace{2mm}\\
  \nonumber
  \centering{\& \hspace{2mm} M20 < -1 \hspace{2mm} \& \hspace{2mm} G >
    0.4 \,\,\, ,} 
\label{eqCAS}
\end{eqnarray}

we are able to recover 92\% of the galaxies visually classified
as early-types in the control sample, with a contamination of only 6\%
from the late-type class (mainly bulge dominated spirals).  The
contours in the different panels of Fig.~\ref{casgm} describe the
areas where 90\% of the total galaxy sample lies, showing how the
control sample is a representative random selection of the parent
population. We therefore adopt Eq.~\ref{eqCAS} to classify
early-type galaxies in the full sample, ending up with 1614 late-type
and 427 early-type objects.
\begin{figure*}
\epsscale{1.}
\plotone{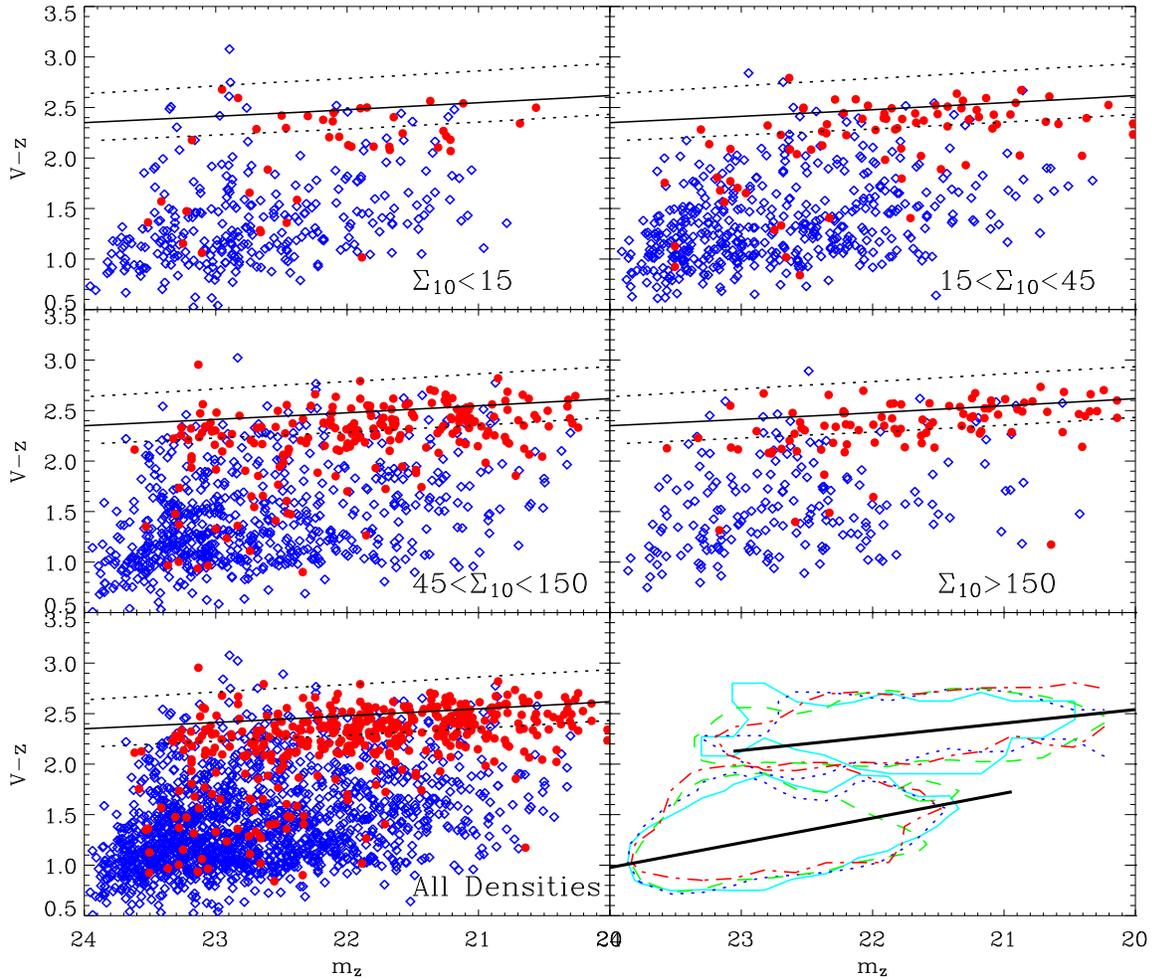}
\caption{ Color-magnitude diagram for galaxies in our sample as a
function of the projected environmental density $\Sigma_{10}$ (in
units of h$^2$ $\ Mpc^{-2}$).  Filled red dots (blue open lozenges)
correspond to objects classified as early-type (late-type) galaxies.
The thin solid line is the red sequence predicted by the
Kodama~\&~Arimoto~(1997) model for a population of early-type galaxies
formed at $z=2$ and observed at $z=0.73$.  The dashed lines correspond
to the same model at $z=0.61$ and $z=0.85$, respectively the lower and
upper limits of the sample.  In the right-bottom panel the iso-density
contours including the 90\% of the red and blue sequences for the 4
environments are reported. The cyan continuous, blue dotted, green
dashed, and the red dot-dashed lines correspond to projected densities
with $\Sigma_{10}<15$, $10<\Sigma_{10}<45$, $45<\Sigma_{10}<150$ and
$\Sigma_{10}>150$, respectively. The thick continuous segments show
the mean slope of the blue and red sequences calculated through
bidimensional gaussian fits to the whole population in the left bottom
panel.  }
\label{col-mag}
\end{figure*}

\section{The color-magnitude relation at $z\sim 0.7$}
\label{cmr}

\begin{figure*}
\epsscale{0.7}
\plotone{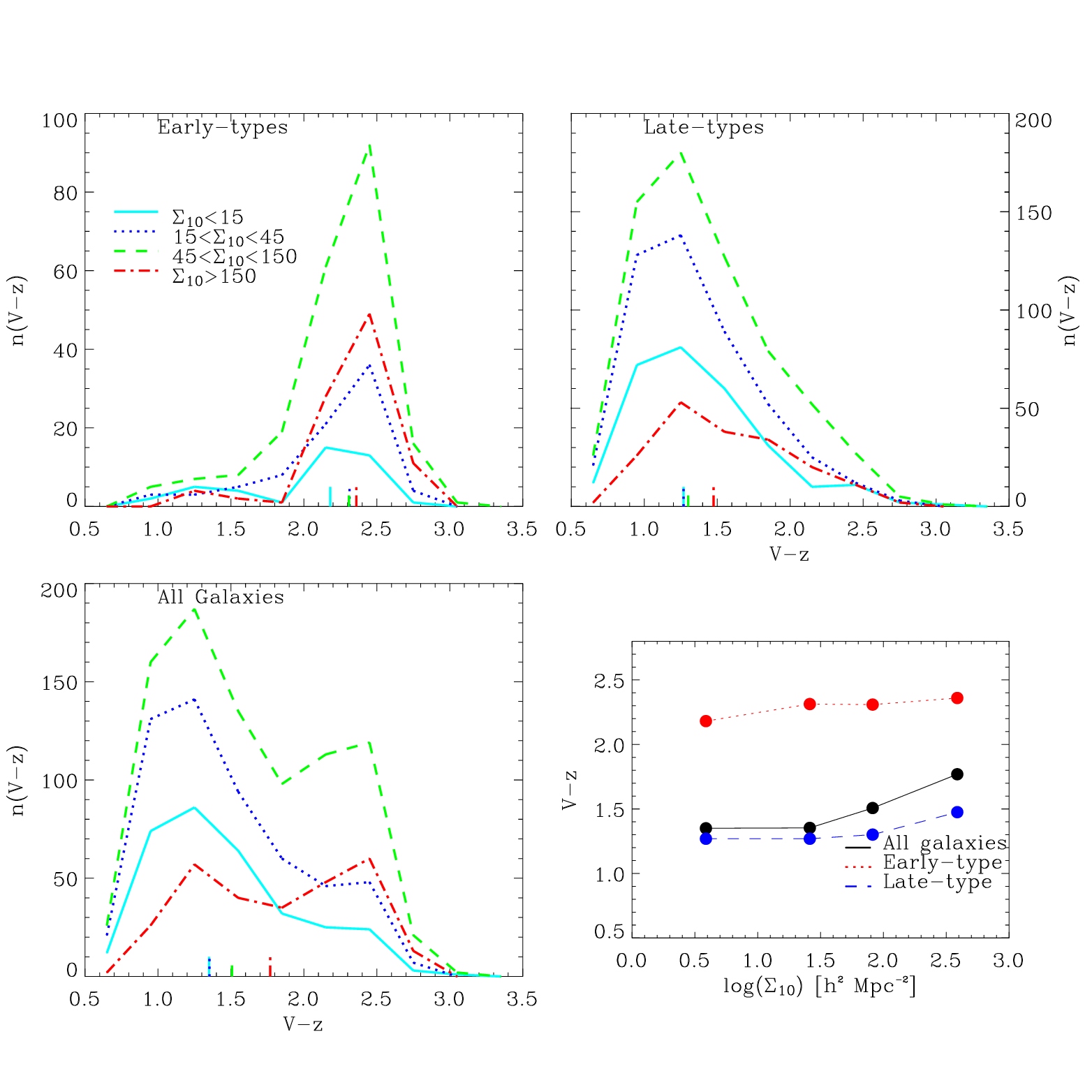}
\caption{The color distribution for galaxies in different environments
and for different morphological types, as indicated in the insert.
The vertical marks on the x-axis in the three panels from bottom-left
to top-right indicate the median of each distribution.  In the
bottom-right panel the relation between the median color and the
projected density is plotted for the entire population
(solid line), for early-type (dotted line) and the
late type galaxies (dashed line).}
\label{col-dens}
\end{figure*}
%
Figure~\ref{col-mag} plots the observed $(V-z')$ color versus $z-$band
magnitude for early- and late-type galaxies in our sample, for
different intervals of local density $\Sigma_{10}$.  Although we are
working with {\it observed} colors, the limited redshift range covered
($z=0.73 \pm 0.12$) makes this plot grossly equivalent to a rest-frame
diagram.  Using observed rather than rest-frame K-corrected colors
reduces the impact of redshift errors and avoids further uncertainties
that are introduced by the SED fitting procedure (see
Franzetti~et~al.~2006). The choice of the $(V-z')$ colour, among the
several available, optimally brackets the 4000 \AA\/ break at
$z=[0.61,0.85]$, maximizing the color contrast of red galaxies.

Even if some contamination by foreground and background objects not
belonging to our redshift slice is certainly present, because of the
uncertainties in photometric redshifts (see Paper I for detailed
discussion), the bimodal nature of the color distribution for the
whole population is evident. The morphological information
contribuites to separate the two sequences more clearly:
a quite well defined red sequence, composed in large majority by
early-type galaxies and with a small color scatter; and a bluer,
somewhat more dispersed one, dominated by late-type morphologies.  The
blue sequence has a larger scatter and a slightly steeper slope than
that of the red sequence.  However, in comparison to other analyses
based on rest-frame colors (e.g. Bell~et~al.~2004a,
Tanaka~et~al.~2005, Franzetti~et~al.~2006) it is remarkably
well-defined.

Also, as shown by the contour plots in the right bottom panel of
Figure~\ref{col-mag}, the two sequences keep their identity
independent of density.
Bi-dimensional Gaussian fits to the two sequences show the following
dependences of the average colors on magnitude:
\begin{equation}
 \left<V-z'\right>_{(red)} = 5.25 - 0.135 m_{z'}
\label{eq1}
\end{equation}
for the red and
\begin{equation}
\left<V-z'\right>_{(blue)} = 7.48 - 0.27 m_{z'}
\label{eq2}
\end{equation}
for the blue sequence. We have verified that these fits do not depend
on the magnitude limit chosen ($m_i=24$), by building a similar plot
for all galaxies brighter than $m_i=25$, and, within $\pm10\%$,
recovering the same best-fit values of eqs.~\ref{eq1} and \ref{eq2}.

Although being dominated by spheroidal and late-type galaxies,
respectively, both red and blue sequences include a mixture of
morphological types.  We will discuss extensively in \S~\ref{A1} the
origin of these two classes.

The solid lines in the Color Magnitude (CM) diagrams of
Figure~\ref{col-mag} show the locus for the red sequence predicted by
the model of Kodama~\&~Arimoto~(1997), for a population of early-type
galaxies formed at $z_f=2$ and observed at $z=0.73$. In this
model galaxies are coeval and described by a single burst of star
formation, after which they evolve passively. The inclination of the
red sequence is produced in this scenario by a metallicity gradient
(see \S~\ref{discussion} for more details). One notes that
while the slope of the data is reproduced fairly well, the model
corresponding to the structure mean redshift seems to be
systematically redder by about 0.1 magnitudes.  We also plot (dotted
lines) the red sequences corresponding to our assumed upper and lower
redshift boundaries, $z=0.61$ and $z=0.85$.  The scatter is comparable
to the region enclosed between the two extreme model sequences.

The effect of the environment on the the color distribution of the
galaxies in this sample is summarized in the histograms of
Figure~\ref{col-dens}, corresponding to the ordinate-axis marginal
distributions of the 2D plots of Figure~\ref{col-mag} (see also
Kodama~et~al.~2001 and Tanaka~et~al.~2005, for similar analyses around
clusters at different redshifts).  The upper panels show the color
distributions for the two morphological classes separately, the
bottom-left one for the total sample.  The vertical long tick-marks on
the x-axis with different line styles give the median values for the
corresponding distributions. The color distribution of the
global sample (bottom-left panel) is bimodal in all environments,
although the distribution of galaxies between the two main sequences
does clearly depend on local density, with the red sequence being less
evident in the lowest density bin, as already noticed in the
color-magnitude diagram.  When looking at the color distribution of
early-types only, an interesting excess of blue galaxies,
independently of the environment, can be noted. At the same time, the
color distribution of late-types show a tail of galaxies with $V-z>$2.
For the rest, the histograms confirm that the color distributions of
spheroidal galaxies, as well as that of late-types, are only weakly
dependent on the environment: the shape of the histograms for the two
classes separately are not sensitive to local density, with median
values rather close to each other, and only a slight tendency for
having redder colors at the highest projected densities.  This is
summarized in the bottom right panel of Figure~\ref{col-dens}, that
reports the median color as a function of local density for the whole
population and for the early- and late-types separately.  When
considered as a whole, galaxies display a clear direct dependence of
the median color on local density, in agreement with the color-density
relation recently observed also at these redshifts by
Cucciati~et~al.~(2006) and Cooper~et~al.~(2006a).  Our analysis
specifically shows how this relation behaves for morphologically
homogeneous classes.  In fact, the median colors of early- and
late-type galaxies separately do not seem to depend strongly on local
density.  We conclude that the bulk of the observed color-density
relation is a by-product of the varying fraction of the two basic
morphological types with environment: the median color for each class
alone shows a mild dependence on local density, but when they are
combined, the median color is dominated by blue spirals/irregulars at
low densities and by red ellipticals/S0s at high densities, thus
producing the observed overall effect.

\section{The morphological mix of the red and blue sequences at $z\sim
0.7$}
\label{A1}
\begin{figure*}
\plotone{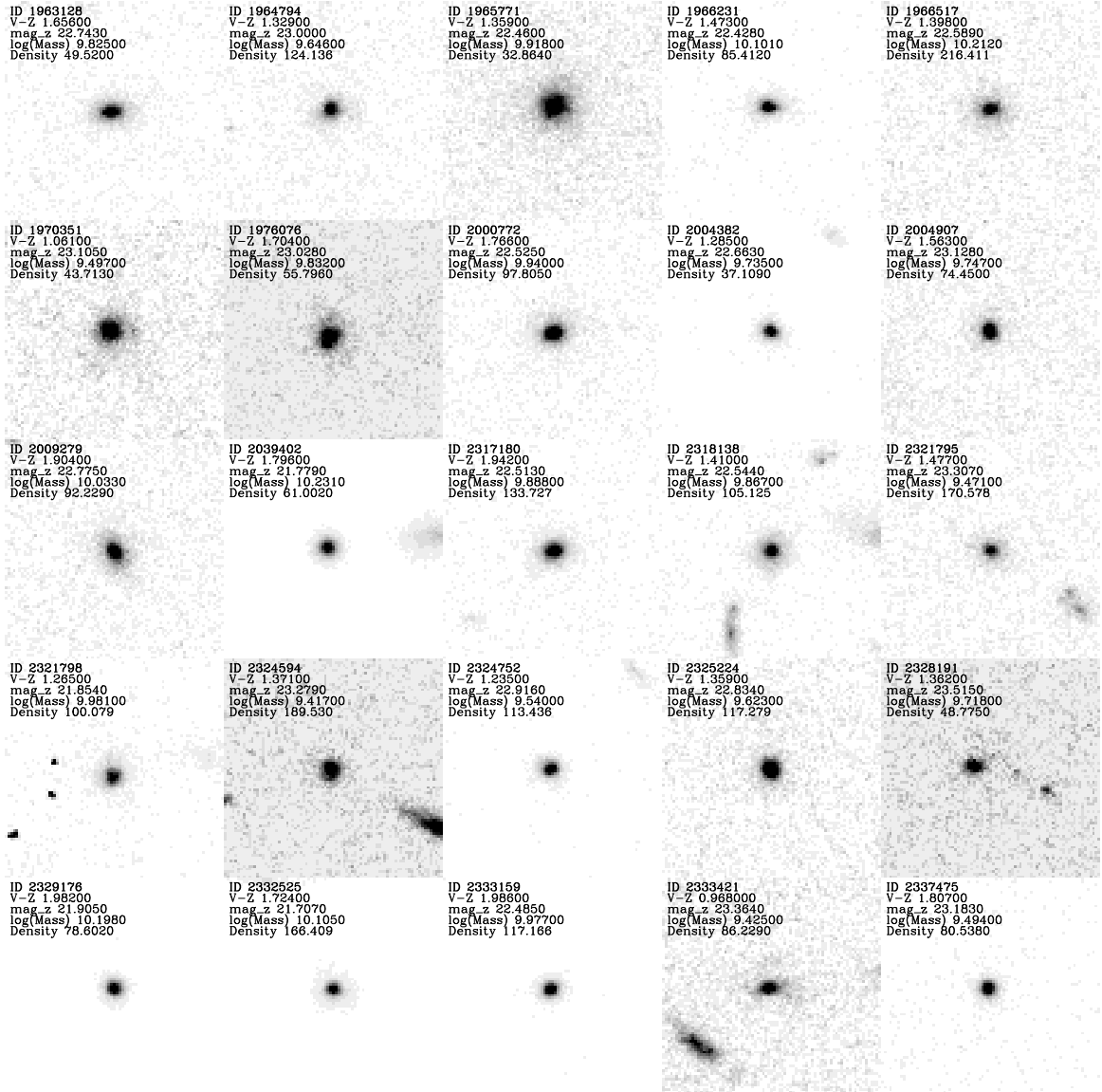}
\caption{ A visual gallery of galaxies in our sample morphologically
classified as early-type, but having blue colors ($V-z'<2$).  For
display reasons, we show only 25 galaxies, randomly selected.
Each postage stamp is 4x4 arcsec, corresponding at $z=0.73$ to
about $20\times 20$ h$^{-1}$ Kpc. In each panel we report the $V-z'$
color, the $z$ magnitude, the stellar Mass and the local
projected density are reported. The bulk of them are clearly
elliptical/S0 galaxies, or in any case characterized by a very compact
morphology.  }
\label{mosaic1}
\end{figure*}
The objects in the color-magnitude plot of Figs.~\ref{col-mag} with
``anomalous'' colors and morphologies, either ``blue early-types'' or
``red late-types'', bear a potential for understanding the
relationship of the two galaxy categories and whether they might
evolve into each other.
Moreover, it is important to quantify the number of late-type
populating the red sequence. When in fact a morphological analysis is
unavailable, due to the lack of high resolution imaging, it is common
to use a simple color criterion to segregate passive early-type galaxies from
star-forming late-type galaxies (e.g. Bell~et~al.~2004a).

\begin{figure*}
\plotone{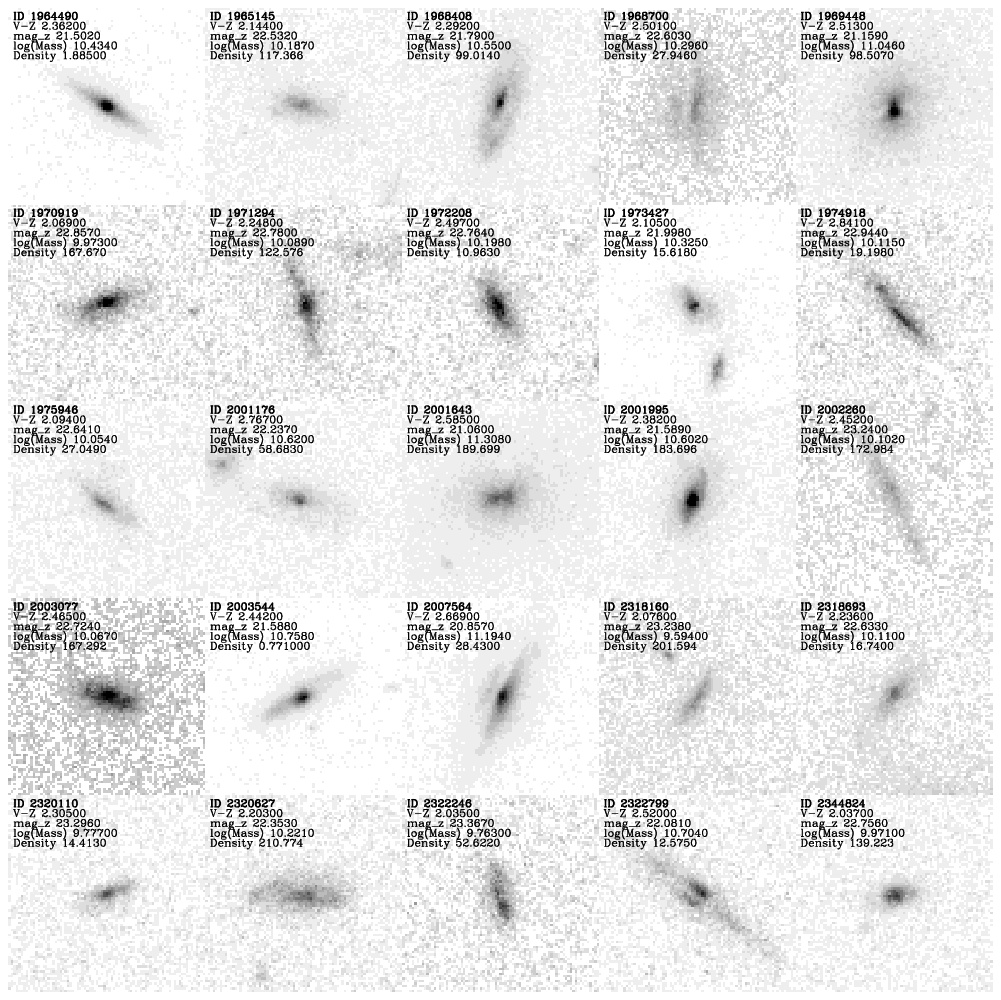}
\caption{ 
Similar to Figure~\ref{mosaic1}, now showing a random sample of 25
galaxies that were morphologically classified as late-type, but
with red colors ($V-z'>2$) that place them on the red sequence.  
Each postage stamp is 4x4 arcsec, i.e. $20\times 20$ h$^{-1}$ Kpc at
$z=0.73$.   Size and information on each panel are as
in Figure~\ref{mosaic1}.  This class is clearly dominated by late
spirals viewed edge-on. The remaining objects include some 
irregular galaxies (probably also dust-reddened), plus a few mergers and
bulge-dominated spirals.} 
\label{mosaic2}
\end{figure*}

We have 78 early- and 184 late-type objects with anomalous colors in
our sample.  The nature of these galaxies is unveiled in
Figure~\ref{mosaic1} and Figure~\ref{mosaic2}, where we have collected
postage-stamp ACS images for 25 early-type-classified galaxies
with ``blue'' colors ($V-z'<2$) and 25 late-type galaxies with ``red''
colors ($V-z'>2$), randomly drawn from the total population in each
class.  For each galaxy, we report the values of the mass, $(V-z')$
color and local projected density $\Sigma_{10}$.  The first
encouraging result from these figures is that the large majority of
the displayed cases are not failures of our automatic morphological
classification.  In fact, in Figure~\ref{mosaic1} all objects are
morphologically dominated by an evident spheroid, while in
Figure~\ref{mosaic2} the prevalent morphology is disk-dominated or
irregular.
%

\subsection{The red late-type population}
\begin{figure}
\epsscale{1.1}
\plotone{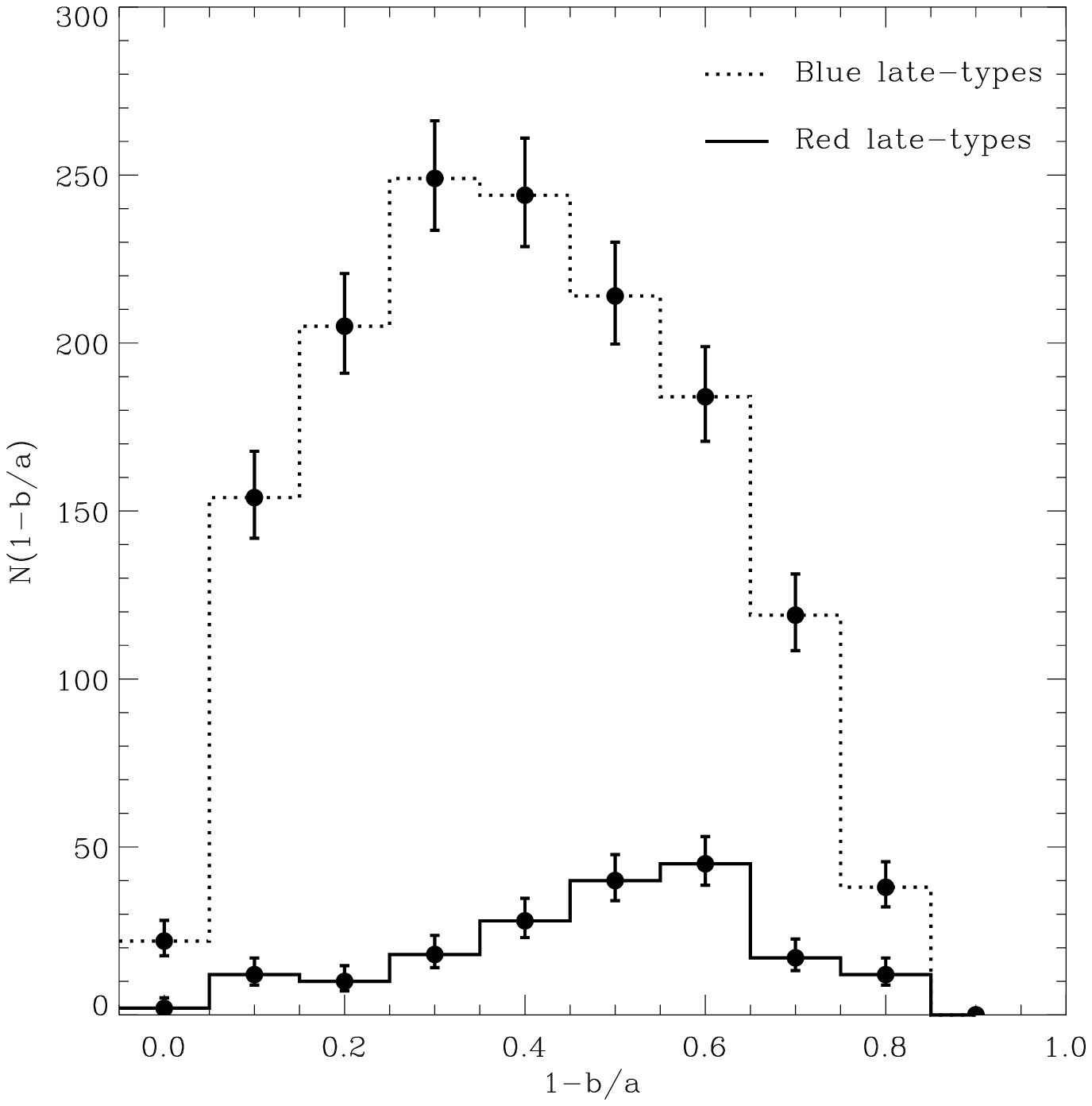}
\caption{Ellipticity distribution for late-type galaxies with $V-z'<2$
and $V-z'>2$ (respectively dotted and continuous line). Error bars
give Poisson noise uncertainties. According to a classical
Kolmogorov-Smirnov test, the probability that the two histograms are
drawn from the same intrinsic distribution is less than $10^{-9}$.}
\label{dist_ell}
\end{figure}
As evident from Figure~\ref{mosaic2}, virtually all the red late-type
objects are edge-on or very inclined spirals, showing red colors
because of strong internal reddening due to their geometrical
orientation.  Weiner~et~al.~(2005), studying the relation between
colors and morphologies for galaxies observed in the DEEP1 survey,
also find a significant number of late-type galaxies showing red
rest-frame colors. They claim in particular that only 75\% of red
galaxies are ellipticals or bulge dominated galaxies. Similarly,
Bell~et~al.~(2004b), using the GEMS survey, find that at $z\sim0.7$
80\% of the $B$-band luminosity density in red galaxies is
contributed by ellipticals or bulge-dominated
galaxies. Franzetti~et~al.~(2006), via spectroscopic measurements,
claim that 35\% of the red galaxies in the VVDS survey show
emission lines due to star formation (with a very small contamination
by AGN).  As in our case, the majority of those galaxies in DEEP1 and
GEMS are edge-on spirals, in which galaxy colors are reddened by dust
absorption, or bulge-dominated spirals.  

To quantify the excess of highly inclined disks observed in
Fig.~\ref{mosaic2}, and to verify the hypothesis that a particular
geometrical orientation of spiral galaxies can explain their extreme
red colors, we studied the distribution of the ellipticities $1-b/a$
for late-type galaxies as a function of their color.  The axial ratio
$b/a$ is related to the galaxy disk inclination $i$ (the angle
between the perpendicular to the plane of the disk and the plane of
the sky) by the Holmberg formula:
\begin{equation}
b/a=\sqrt{\left(1-q_0^2\right)\left(\cos^2i\right)+q_0^2},
\end{equation}
where $q_0$ is the intrinsic ratio between the height and the radius
of the disk (a typical value for $q_0$ is 0.2,
Tully~\&~Pierce~2000). In Fig.~\ref{dist_ell} we compare the
distribution of ellipticities that we have measured for late-type
galaxies lying in the blue and red sequences. The dotted line, in
particular, shows the ellipticity for blue late-type galaxies: the
excess of intermediate ellipticity values ($1-b/a\sim0.3$) is very
probably due to the large number of irregular/merging objects, that
are mixed to spiral galaxies in the late-type class.  For red
galaxies, instead, the distribution is significantly skewed towards
higher ellipticity values, peaking at $1-b/a\sim0.8$. A
Kolmogorov-Smirnov test shows that the probability that the two
ellipticity distributions are drawn from the same intrinsic population
is less than $10^{-9}$.  This confirms that the inclination of the
disk is the dominant effect in determining the measured color for
late-types that are observed well redward of the blue sequence.

Consequently, a simple color criterion to identify passive elliptical
galaxies will always suffer of a certain degree of
contamination. Based on our 184 ``red spirals'' out of 533 galaxies,
this contamination amounts to $\sim $33\%. These red late-type galaxies
contribute to 20\% of the luminosity density on the red sequence, a
figure that confirms the result by Bell~et~al.~(2004b),
Weiner~et~al.~(2005) and Franzetti~et~al.~(2006).  Remarkably, our
percentage of red galaxies with an early-type morphology is comparable
to that measured in the local Universe for SDSS galaxies (58\%, see
Renzini~2006).  Also, a parallel study of the evolution of the 
global early-type population in the COSMOS field
(Scarlata~et~al. 2007) is in very good agreement with our estimate at
$z\sim 0.7$, showing also a similar fraction between $z=0.2$ and $z=1$. 
Taking into account the differences in the selection criteria,
these results indicate that the fraction of red galaxies
with a relaxed morphology remains substantially unchanged between
$z=1$ and the current epoch.

\subsection{The blue early-type population}
\label{orig-bimod}

Whereas the red late-type galaxy population is clearly the
result of a geometrical effect, the colors of the {\it bonafide} blue
early-types could be indicative of recent or ongoing star-formation
activity, placing them outside of the red sequence.  These are
interesting objects, as they may represent a transition stage, 
in the process of migrating onto the red sequence.

Evidence for blue colors or star-formation signatures in the spectra
of early-type galaxies has been reported by various studies in the
past (Schade~et~al.~1999; Im~et~al.~2001; Menanteau, Abraham \& Ellis
2001; Menanteau~et~al.~2004; Cross~et~al.~2004; Cassata~et~al.~2005;
Treu et al. 2005a; Ilbert~et~al.~2006).  Schade~et~al.~(1999), in
their study of CFRS galaxies find a percentage of about 30\% of
morphologically- selected spheroidals that exhibit [OII]$\lambda3727$
emission.  Cassata~et~al.~(2005), studying the morphological
properties of the $K20$ sample (Cimatti~et~al.~2002), recognize a
class of "peculiar ellipticals", i.e.  spheroidal galaxies with some
distortion of the isophotes or with signs of a recent interaction. The
majority of these galaxies show evidence for recent star formation in
their spectra.  The morphologies of the blue early-type galaxies found
here are quite compatible with these properties.  A direct check of
their spectroscopic properties will soon become possible thanks to the
ongoing z-COSMOS redshift survey at the VLT (Lilly~et~al.~2007).

In our sample, we have 78 blue objects among the 427 spheroidals
($\sim$18\%). This percentage is lower than that
  measured at similar redshifts by various authors (Im~et~al.~2001,
  Cross~et~al.~2004, Ilbert~et~al.~2001), who indicate fractions around
30\%. This is not surprising, as
  our sample is centred on a large-scale structure that is clearly
  over-abundant in high-density regions, including the very massive
  X-ray/lensing cluster (Guzzo et al. 2007), making it possibly richer
  in early-type objects than an average volume at the same redshift.
The blue early-types concur to 5\% of the mass in ellipticals and to
less than 3\% of the total mass in galaxies. This means that if these
will eventually migrate to the red sequence, they are not able to
increase significantly its mass content.

\begin{figure}
\epsscale{1.1} \plotone{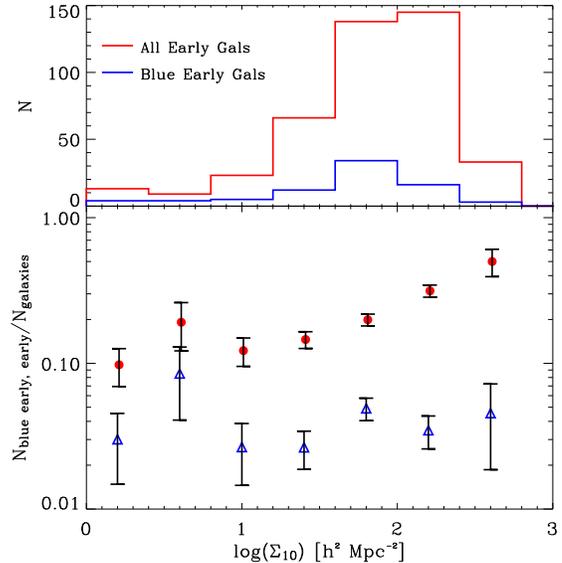}
\caption{ {\it Top}: The distribution of ``blue early-type'' galaxies
as a function of local density $\sigma_{10}$ (dotted blue line),
compared to that of the whole population of early-type galaxies
(continuous red line).  {\it Bottom}: The corresponding relative
fractions (triangles and filled circles, respectively).  While the
global population of early-type galaxies shows clearly the familiar
Morphology-Density relation (see Paper I), the blue early-type
fraction is substantially independent of local density. }
\label{frac-col}
\end{figure}

Figure~\ref{frac-col} investigates a possible relation between the blue
early-type population and the environment. Specifically, we plot the
fraction of objects as a function of local density separately for all
early-type galaxies and for those with blue colours only.  While the
fraction of all early-types rises with local density, (see Paper I for
a full discussion of the morphology-density relation at this redshift)
the percentage of blue spheroidals remains constant as a function of
the environment, about $2-5\%$. A Kolmogorov-Smirnov test confirms
that the two distributions are statistically distinct, as the probability
that they are drawn from the same original population is less than $10^{-9}$.

Could these objects be precursors of some of today's red early-type
galaxies?  Ferreras~et~al.~(2005) suggest that a robust criterion to
isolate the progenitors of early-type galaxies at different redshifts
should exclude those objects that, even though showing a clear
elliptical morphology, will not plausibly evolve into the local
Kormendy Relation at the current epoch (see also
Scarlata~et~al.~2007).  The Kormendy Relation relates the effective
radius $R_{50}$ (the radius, in physical units, containing half of the
total light) with $\mu(<r_{50})$, the mean surface brightness within
$r_{50}$ (in arcsec).  In this plane, local bright elliptical galaxies
show a relatively tight scaling relation, for sizes between $\sim$1
and 100 Kpc (Kormendy~1977). This is nothing else than a projection of
the well-known fundamental plane of elliptical galaxies, connecting
these two quantities to the galaxy velocity dispersion.
Capaccioli~et~al.~(1992)  show that more in general
galaxies form 2 distinct families in the $\mu(<r_{50})$ vs $R_{50}$
plane: while luminous ellipticals follow the classical {\it Kormendy
relation}, some S0s, spiral bulges and dwarf ellipticals populate a
narrow nearly vertical strip with sizes typically smaller than
$R_{50}\lesssim3$ Kpc and a wide range of surface brightnesses.

\begin{figure*}
\epsscale{1.}
\plotone{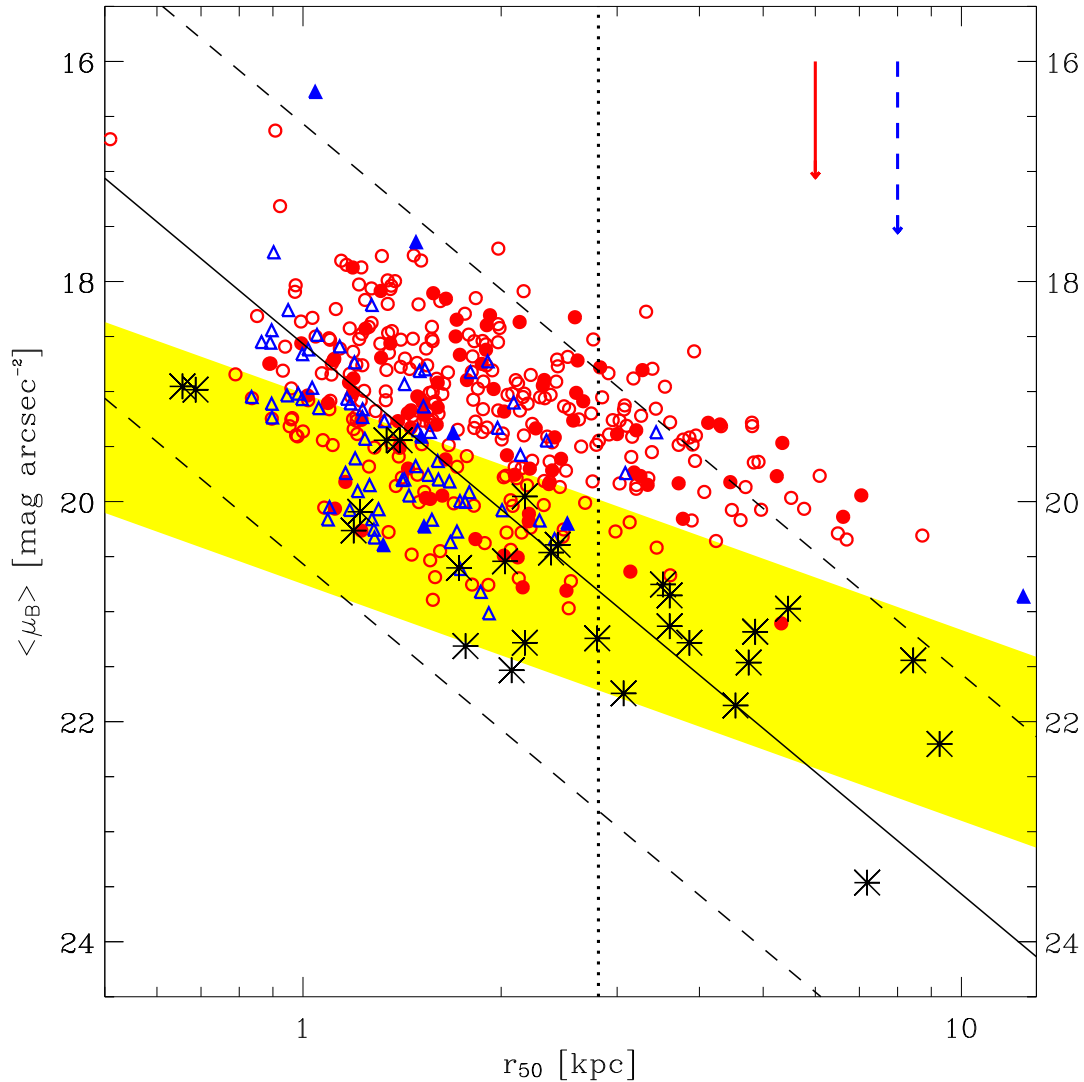}
\caption{ The Kormendy relation for the 427 early-type galaxies in the
sample. Triangles and circles indicate respectively blue and red
early-type galaxies as defined in the text, with filled symbols
corresponding to objects lying in high-density regions
($\Sigma_{10}>150$ h$^2$ Mpc$^{-2}$) . The shaded band
indicates the $2-\sigma$ corridor described by local Coma Cluster
galaxies (reported as asterisks from Jorgensen~et~al.~1995).
The continuous line shows the locus of constant absolute magnitude
$M_B$=-20, with the dashed lines corresponding to $\Delta M_B=\pm2$
around this.  The vertical line at $R_{50}$=2.81 indicates the upper
size limit for the ``ordinary group'' in Capaccioli~et~al.~(1992),
that is composed mainly by dwarf ellipticals, S0 and bulges of
spirals.  The vertical continuous and dashed arrows show the
maximum fading expected between $z=0.73$ and 0, respectively for
passively-evolving and ``maximally star-forming'' early-type galaxies
(see text). Note that here efective radii are given for $h=0.7$. }
\label{kormendy}
\end{figure*}

We have therefore computed the surface brightness in the $B$
rest-frame band, $\mu_B$, and the corresponding $R_{50}$ from the
ACS/F814W band, for the early-type galaxies in our sample.  The result
is plotted in Fig.~\ref{kormendy}.  To obtain the plotted quantities,
one has to take into account that due to the cosmological dimming the
observed surface brightness $\mu_{\nu}$ is related to the rest-frame
intrinsic surface brightness $\mu_{\nu,0}$ as
\begin{eqnarray}
\mu_{\nu}=\mu_{\nu,0}+10\log(1+z)+K(\nu,z)\\
K(\nu,z)=-2.5\log\left[\frac{L_{\nu(1+z)}}{L_{\nu}}(1+z)\right]
\end{eqnarray}
where $\mu_{\nu,0}$ is the rest-frame surface brightness, $(1+z)$
reproduces the redshifting of the bandwidth, the ratio between
$L_{\nu(1+z)}$ and $L_{\nu}$ accounts for the difference in flux
between the observed and emitted bands, and $K(\nu,z)$ is the $K$-correction.
For the redshift range we are exploring ($z=[0.61,0.85]$), the
observed $F814W$ band well matches the rest-frame $B$ band.  Hence, to
first approximation, $L_{\nu(1+z)}\cong L_{\nu}$, and the
$K$-correction reduces to $K=-2.5\log(1+z)$, without any evolutionary
correction. The main uncertanties on the measured quantities are due
to the scatter in the photometric redshifts, that affect the
determination of distances and the surface brightness dimming
correction. We estimate a typical error of less than 10\% on $R_{50}$,
and $\sim$0.2 mags on $\mu_B$.  For comparison, we also plot in
Fig.~\ref{kormendy} the points corresponding to early-type galaxies in
the Coma Cluster (Jorgensen~et~al.~1995), together with the $2\times
rms$ corridor around the best fit (grey/yellow band).

We can notice that, in our sample, blue early-type galaxies have
typically $R_{50}<3$Kpc, while early-type galaxies larger than this
size have normally red colors. Also, for a fixed $R_{50}$ blue
galaxies have fainter $\mu_B$ than red ones.  On the contrary,
the most luminous red galaxies follow a flatter relation, virtually
parallel to the $z=0$ Kormendy relation, with sizes in the range
3 Kpc $<R_{50}<10$ Kpc. Thus, blue elliptical galaxies seem to form a
different population with respect to the most luminous red ones.

Simple evolutionary considerations can be used to establish which of
the blue and red galaxies can evolve onto the local Kormendy Relation,
and can thus be considered progenitors of nowadays spheroidals. Purely
passive evolution models (Bruzual~\&~Charlot~2003) with an initial
burst of 1 Gyr and a Salpeter IMF, predict a dimming of 0.7-1
magnitudes between $z=0.7$ and $z=0$, for a redshift of formation
between 2 and 5.  If this is the case, the majority of our red
early-type objects will move onto the yellow/grey strip indicating the
local relation for Coma elliptical galaxies.  It seems thus reasonable
to conclude that the large majority of our red early-type galaxies
could well be progenitors of local ellipticals, in terms of their
surface-brightness/radius properties.  We should also consider that
the $z\sim0$ Kormendy relation shown in Fig.~\ref{kormendy} strictly
holds for early-type galaxies in high density environments, being
based on Coma-cluster galaxies. In our sample, instead, only $\sim90$
elliptical galaxies among 427 live in high density regions
($\Sigma_{10}>150$).  These galaxies are plotted as filled circles in
Fig.~\ref{kormendy}, showing that they do not seem to differ from the
overall population of red early-types in this diagram.

A fading of 0.7 to 1 magnitude is actually a lower limit to the
luminosity decrease of any galaxy between $z=0.7$ to $z=0$ (see also
Gebhardt~et~al.~2003, Van~Dokkum~\&~Stanford~2003,
Treu~et~al.~2005b). Blue objects will in general fade by an even
larger amount. A reasonable estimate can be obtained by a
model which continuously forms stars at a constant rate from the
redshift of formation, down to the redshift of observation ($z=0.73$
in our case); at that point star formation is switched off.
Implementing this recipe using the Bruzual~\&~Charlot~(2003) code and
considering formation redshifts between 2 and 5, we obtain an expected
fading of 2.5 magnitudes between z=0.73 and today.  The decay in
magnitude will be even larger than this if these galaxies are
experiencing a short-lived star-burst triggered for example by an
interaction/merger.

Even conceiving a fading as small as 1.5 magnitudes for our blue
sample (which is unrealistic if these galaxies are starbursting),
then, just the 26 objects lying above the yellow/grey region
(i.e. $\sim 30\%$ of the total sample of blue spheroidals), would end
up onto the local Kormendy relation. More in general, those with surface
brightness fainter than $\sim$19.5 mag~arcsec$^{-2}$ should eventually
move below the local relation, occupying the region with
$R_{50}<3$Kpc where Capaccioli~et~al.~(1992) find mainly bulges of
spirals, S0 galaxies and dwarf spheroidals.  Could our blue
spheroidals be a mixture of these three classes?

One one side, looking carefully at the morphologies shown in
Fig.~\ref{mosaic1} (and to the other galaxies in the same sub-sample),
we see that only about 10\% of the objects show signs for a faint disk
component. We cannot exclude that, additionally, some very low-surface
brightness disks are in fact below the SB limit of the ACS data.
However, previous studies based on deeper, multi-band HST imaging (as
GOODS, Cassata~et~al.~2004) do not find evidence for a general
population of low-surface brightness disks around similar
objects. Also, resolved color analysis of similar objects shows
evidence of blue cores rather than blue disks
(Menanteau~et~al.~2004). At the same time it seems also difficult to
think of these objects as precursors of S0's, for the very simple
reason that we do not detect any correlation with density of their
relative fraction, that is instead observed locally for S0
(e.g. Dressler 1980).  Finally, if the fading is comparable to that
computed from our simplified star-forming model (i.e. of the order of
2.5 magnitudes), we see that only a fraction of these objects might
become as faint as to be classified as dwarf ellipticals.  Considering
a blue elliptical in our sample with $\mu_B$=20 and $R_{50}$=1.5 kpc,
this implies a total magnitude $m_B$=21.4 and thus an absolute
magnitude $M_B = -21.6$.  Assuming as a definition of a dwarf
ellipticals a magnitude limit $M_B>$-18, we see that a fading of more
than 3.5 magnitudes would be required.  This is not implausible, but
would require that our blue spheroidals are typically experimenting a
strong burst of star formation.

\section{Understanding the CM diagram: the color-mass relationship }
\label{cMassr}

\begin{figure*}
\epsscale{1}
\plotone{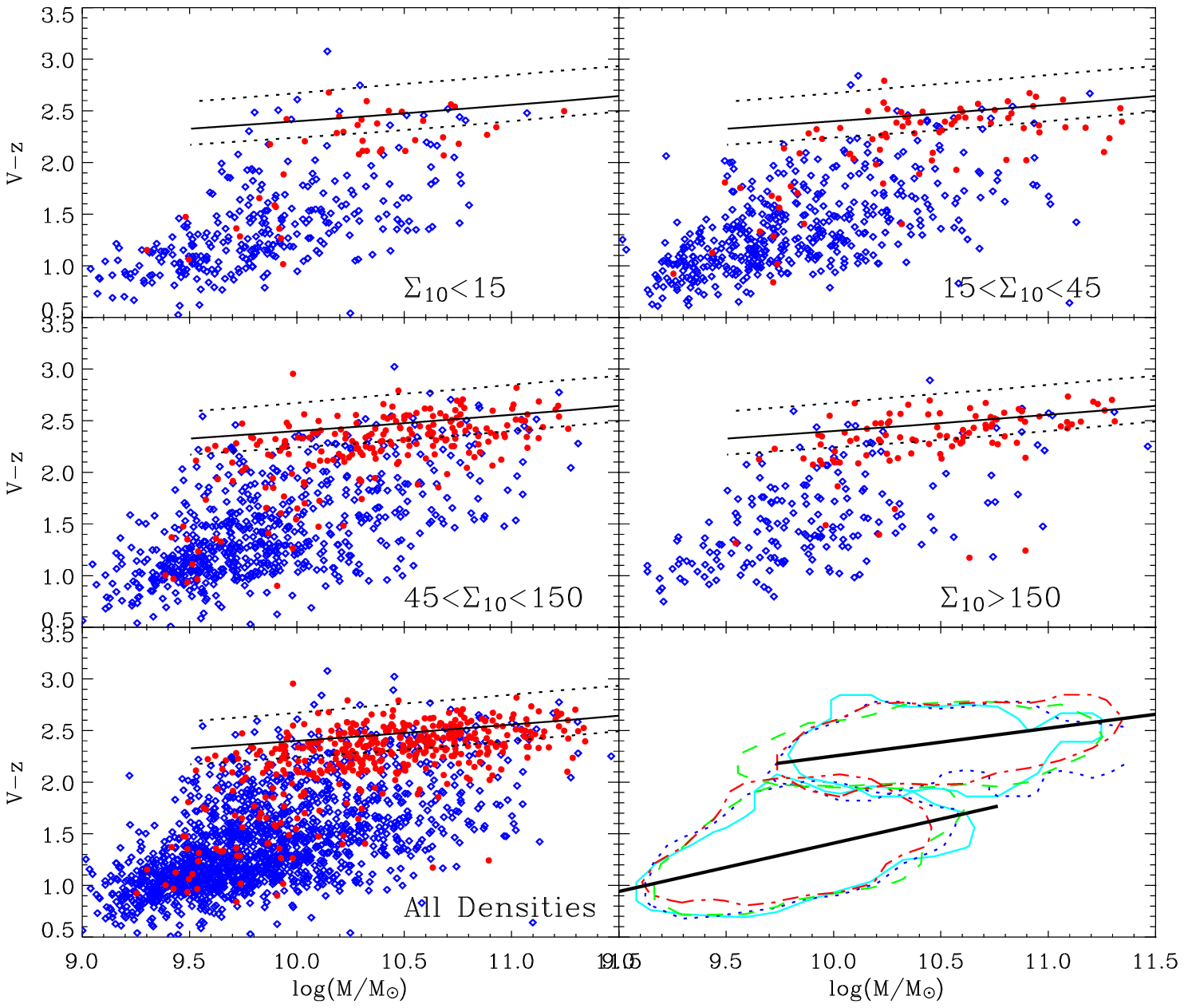}
\caption{ Relationship between color and stellar masses as a function
of the projected environmental density $\Sigma_{10}$ (in units of
h$^2$ $\ Mpc^{-2}$).  Filled red dots (blue open lozenges) correspond
to objects morphologically classified as early-type (late-type)
galaxies.  The thin solid lines are the red sequence predicted by the
Kodama~\&~Arimoto~(1997) model, thin dashed lines the same model
calculated for $z=0.61$ and $z=0.85$.  In the right bottom panel the
contours of the red and blue sequences in the 4 environmental regimes
are reported.  The cyan continuous, blue dotted, green dashed, and the
red dot-dashed lines correspond to projected densities with
$\Sigma_{10}<15$, $15<\Sigma_{10}<45$, $45<\Sigma_{10}<150$ and
$\Sigma_{10}>150$, respectively. The thick continuous segments show
the mean slope of the blue and red sequences from bi-dimensional
gaussian fits to the whole population.  }
\label{col-mass}
\end{figure*}
%
It is interesting to recast the CM diagrams of Figure~\ref{col-mag} in
terms of stellar mass.  This is shown in Figure~\ref{col-mass}.  Also
in this case, two well-defined red and blue sequences, respectively
dominated by spheroidal and late-type galaxies, are evident.  The red
sequence is more clearly defined as the projected density increases,
whereas in the lowest density bin it is nearly absent, similarly to
the CM plot.  With this noted exception, in all other density regimes
the red and blue sequences, again, keep remarkably invariant.  The
bottom-right panel summarizes these results by plotting the density
contour plots in the Color-Mass plane for the red early-type and blue
late-type galaxies (that is galaxies that satisfy both morphological
and color criteria) in various density regimes.  The remarkable
consistency of the contours confirms the substantial indipendence of
the color-mass relation {\it within each morphological class} on the
environment.  The thick continuous lines correspond to the average
dependences of red and blue galaxies found with a 2D gaussian fitting.
While the slopes of the two sequences in the CM diagrams of
Figure~\ref{col-mag} were fairly similar (just slightly steeper in
eq.~\ref{eq2} than in eq.~\ref{eq1}), in the current plots the blue
sequence is steeper than the red one:
\begin{equation}
\left<V-z'\right>_{(red)} \simeq -0.15 + 0.245 \log (M/M_\odot) , 
\label{eq3}
\end{equation}
\begin{equation}
\left<V-z'\right>_{(blue)} \simeq -3.4+ 0.47 \log (M/M_\odot)  .
\label{eq4}
\end{equation}
We have tested (as for Eq.~\ref{eq1} and \ref{eq2}) that these fits do
not change by including galaxies with $24<m_i<25$. 

\section{DISCUSSION}
\label{discussion}

In this paper we have detailed the morphological composition of the
blue and red parts of the CM diagram at $z\sim 0.7$ as a function of
the environment, extending previous results from combined HST -
ground-based surveys (Bell~et~al.~2004b, Weiner~et~al.~2005). Our
results show at $z\sim0.7$ the substantial invariance of the main
features in this diagram of galaxies with respect to local density,
similarly to what was found by Hogg~et~al.~(2004) at $z\sim0$. The
properties of the red sequence at moderate/high redshifts have so far
been so far studied in particular in clusters (see
e.g. Gladders~et~al.~1998, Kodama~et~al.~2001, Blakeslee~et~al.~2003,
Yee~et~al.~2005, Tanaka~et~al.~2005), or using rest-frame colors, for
large comprehensive galaxy surveys (as SDSS, e.g.  Hogg~et~al.~2004,
Combo-17, Bell~et~al.~2004a, DEEP-2, Cooper et al. 2006a,b and
VVDS, Cucciati et al. 2006 and Franzetti~et~al.~2006).  On the
other hand, the evidence for a ``blue sequence'' when considering
galaxy colors at a fixed redshift emerges here more clearly than in
previous works.  As it is evident from Figure~\ref{col-mag}, the
separation in morphological classes helps to better separate the two
color populations.  These patterns become only slightly more confused
confused when other colors, not bracketing the 4000$\AA$ break, are
used instead of the $V-z'$.  This suggests that they are not simply
the result of a population-dependent Balmer jump, but derive from more
general properties of the UV-optical spectrum.

We have shown in Fig.~\ref{dist_ell} that a significant fraction of
highly inclined spiral galaxies are observed to have red colors only
because of internal extinction. Correcting for this would clearly have
the effect to move them down in the CM diagram.  When photometrically
estimated stellar masses are plotted instead of galaxy luminosities
(Figure~\ref{col-mass}), the galaxy distribution patterns keep some
similarities, but also show differences, compared to the CM plots.
In particular, the two red and blue color sequences show steeper
dependences on galaxy mass (eqs. \ref{eq3} and \ref{eq4}) than on
magnitude.  Another difference is that, while the slopes of the red
and blue sequences in the CM plot appear to be similar (cf. bottom
panel of Figure~\ref{col-mag}) , the blue color-mass sequence is
significantly steeper than the red one.  Figure~\ref{col-mass} shows
that galaxies with colors $V-z'\sim 0.5$ to $\sim 1.7$ and masses in
the range $10^9$ to $\sim1-3 \times 10^{10}\ M_\odot$ are dominated by
late-type morphologies, while the locus with colors from $V-z'\sim 2$
to $\sim 3$ is dominated by spheroidal systems with $\sim 10^{10} < M
< 10^{11.5}\ M_\odot$. These transition scales in mass are comparable
to those shown at z$\sim$0 by Scodeggio~et~al.~(2002) and
Kauffmann~et~al.~(2004) and agree with the trend of an evolving star
formation quenching mass suggested by Bundy~et~al.~(2006).

The models by Kodama~\&~Arimoto~(1997), that we have superimposed on
top of our observed red sequence distributions in
Figures~\ref{col-mag} and \ref{col-mass}, have been originally
developed to explain the tight red sequence observed in galaxy
clusters. In these models, galaxies are co-eval (with formation
redshift $z_f=2$ in our case\footnote{We also tested models with
$z_f=5$, that give, for the same redshift, a slightly redder
sequence.}) and are described by a classical single burst of star
formation and subsequent passive evolution in color and luminosity
down to the time of observed galaxy redshift. The slope of the red
sequence is explained as a differential metallicity effect: galaxies
with smaller mass have more difficulty to retain the metals ejected by
supernova explosions, and thus look bluer than their more massive
companions with the same age.  On the other hand, the blue/late-type
sequence shows a significantly steeper dependence of color on mass,
and cannot be explained in terms of metallicity, especially at such
blue colors.  It seems more physically plausible to associate the
color of blue galaxies to the timescale of star formation
(Searle,~Sargent~\&~Bagnuolo~1973, Larson~\&~Tinsley 1978). Thus, the
observed blue sequence can be interpreted as a relationship between
the actual specific star formation, evidenced by the galaxy color, and
its mass.  In this picture, the observed steep sequence would indicate
a gradual fading of star formation, with bluer hot stars gradually
disappearing for late-type galaxies of larger and larger mass.  This
shows in quite an explicit way, at a fixed epoch, the so-called
``downsizing'' effect, now observed in several galaxy surveys
(e.g. Cowie~et~al.~1996, Gavazzi~et~al.~1996,
Franceschini~et~al.~1998, Treu~et~al.~2005a, Bundy~et~al.~2006).  This
picture is reinforced by the observation that the slope of the blue
sequence in the color-mass plane does not depend on the value of
the local density (Figure~\ref{col-mass}), indicating that the
specific amount of young stars produced per unit mass in late-type
galaxies does not depend on the environment and is essentially
regulated by internal processes.

What is dependent on the environment is the relative number of
early- and late-type galaxies populating the two sequences, i.e. the
well-known morphology-density relation, still present at these
redshifts (see Paper I and references therein).  In this respect, our
results agree very well with those recently obtained at the same
redshift by the VVDS (Cucciati~et~al.~2006) and DEEP-2
(Cooper~et~al.~2006b).  The VVDS work, in
particular, explcitly claims that {\it the color-magnitude
distribution is not universal but strongly depends upon environment},
a statement which might seem in apparent contraddiction with the
findings of our work.  This apparent confusion needs to be clarified.
What this sentence and the overall Cucciati et al. work imply, is
simply that there exist a color-density relation observed at virtually
any redshift (with a tendency to disappear and possibly even invert
its trend for $z>1.2$, which is the main result of that paper), and
thus that the density of points occupying the red and blue parts of
the diagram is a function of environment.  Again, this is simply the
color-density or morphology-density relation that we observe also
here.  What we mean here by stating that {\it the color sequences do
not depend on the environment}, is that the location and slope of the
two main clouds is independent of the local density, as shown by
Fig.~\ref{col-mag}.  This is also implicitly noted by Cucciati et al., that
find that {\it the location of the colour gap between the red and blue
peaks appears to be roughly constant and insensitive to the
environment at all redshifts}.  In their Fig.~10, in fact, while the
blue sequence is too compressed to define a slope, it is very clear
that the red sequence position and slope is rather insensitive, at all
redshifts, to local density.

\section{SUMMARY AND CONCLUSIONS}
\label{end}

The main results we have obtained in this paper can be summarized as
follows.

\begin{itemize}

\item The $\left<V-z'\right>$ color-magnitude diagram at $z\sim 0.7$
for our sub-sample of the COSMOS catalogue shows a clear red sequence
compatible to the expectations of the passive models of Kodama \&
Arimoto~(1997).  Rather than forming a ``blue cloud'', blue galaxies also
tend to aggregate along a similar, bluer sequence with increasing
$\left<V-z'\right>$ average color with luminosity.  The distinction
and ``sharpness'' of the two sequences is enhanced by separating
galaxies into morphological types, with early (late) types dominating
the red (blue) sequence.

\item The distribution function of galaxy colors is a function of the
environment, as locally measured by Balogh~et~al.~2004, using
SDSS, with a distribution more skewed towards redder colors in in
high-density regions. This confirms the positive color-density
relation measured at similar redshift by Cucciati~et~al.~(2006) and
Cooper~et~al.~(2006a). However, when early- and late-type galaxies are
separated, the distributions become nearly independent of the
environment.

\item We also study in detail the population of ``anomalous''
galaxies, with contrasting colors and morphologies.  ``Red
late-types'' are found to be mostly edge-on spiral galaxies, for which
colors are dominated by internal reddening by dust. Their ellipticity
distribution in fact is dominated by a peak around $1-b/a\sim0.8$.  In
a simple color-selected sample, these galaxies would represent a 33\%
contamination over the population of truly red, passive
spheroidals. At the same time, red late-type galaxies are less
luminous than red early-types, thus contributing only to the 20\% of
the luminosity density of red galaxies (similarly to
Bell~et~al.~2004b).

\item We also identified an interesting population of ``blue
  spheroidals'', that show a flat Morphology Density relation, at
  variance with that of red early-types.  They are also found to have
  smaller physical half light radii and fainter surface brightnesses,
  and contribute to less than 3\% of the total galaxy mass
  content. Simple considerations about the evolution of their stellar
  populations lead us to conclude that at least 70\% of them will not
  evolve into the $z=0$ Kormendy relation for bright ellipticals, but
  will occupy a region in the $\mu_B-R_{50}$ plane where nowadays we
  find S0 galaxies, spiral bulges and dwarf ellipticals. None of these
  three classes seems to be an obvious candidate for the descendents
  of our blue spheroidals.


\item In a color-mass diagram color sequences are even better defined,
with red galaxies covering in general a wider range of masses at
nearly constant color, and blue galaxies showing a more pronounced
dependence of color on mass.  We interpret the blue sequence as a
specific star formation sequence, thus with the less massive late-type
galaxies showing a younger average age of their stellar population.
Notably, the slope of this sequence has no dependence on the
environment, indicating that the specific amount of young stars
produced per unit mass does not depend on local density, but seems to
be regulated more by the total galaxy mass itself.  What the
environment does, is to regulate the relative number of galaxies
between the red and blue sequences, which at this redshift we still
find (once cleaned of spurious effects as internal extinction due to
inclination in spirals), is nearly synonimous of being morphologically
an early-type or late-type galaxy, respectively.  This could be
obtained via {\it ab initio} processes (e.g. by simply forming more
massive galaxies, which form stars and become redder faster, in
higher-density regions, as it is natural in hierarchical models).
Alternatively, rapid bursts of star formation and/or gas depletion
could be induced later in the life of the galaxy, as e.g. due to
interactions/mergers in groups or ram-pressure stripping during infall
of the galaxy into clusters or cluster-cluster mergers (see evidence
for this in Paper I).  Both classes of mechanisms would seem to be
able to rapidly move the galaxy away from the `quiet life' of the blue
sequence (where things develop driven simply by the galaxy mass) and
place it into the 'passive repository' of the red sequence.  Using the
full-fledged COSMOS data set we hope to be able in the near future to
be able to disentangle the respective role of these mechanisms,
perhaps including also the possibly crucial contribution, not
considered here, of switching-on an AGN in the galaxy nucleus.

\end{itemize}

\acknowledgments

PC thanks the Osservatorio Astronomico di Brera for hospitality during
the development of this work.  LG thanks C. Firmani and A. Boselli for
useful discussions.  We thank T. Kodama for providing us with his
model red sequences in electronic form. PC thanks P. Franzetti for
providing Bruzual \& Charlot models.  We thank Marcella Carollo and
Simon Lilly for suggesting the Kormendy relation analysis.

\end{document}